\documentstyle[12pt]{article}

\newlength{\dinwidth}
\newlength{\dinmargin}
\def\bbbone{{\mathchoice {\rm 1\mskip-4mu l} {\rm 1\mskip-4mu l}
{\rm 1\mskip-4.5mu l} {\rm 1\mskip-5mu l}}}
 \newcommand{\subs}[1]{\mbox{\scriptsize\rm #1}}
 \newcommand{\esk}{\enspace ,}
 \newcommand{\esp}{\enspace .}
 \newcommand{\mcr}{m_{\subs{cr}}^2}
 \newcommand{\p}{$\phantom{-}$}
 \newcommand{\pn}{\phantom{0}}
 \newcommand{\A}{{\cal A}\,}
 \newcommand{\Cstar}{C^{\ast}}
 \newcommand{\Dirac}{\mbox{$\not\!\!D$}}
 \newcommand{\Dm}{\triangle m^2}
 \newcommand{\Nc}{N_{\subs{c}}}
 \renewcommand{\S}{{\cal S}\,}

\begin{document}

 \noindent{\tt DESY $93-046$ \hfill ISSN $0418-9833$}\\
 {\tt April 1993}
 \renewcommand{\thefootnote}{{\protect\fnsymbol{footnote}}}

 \begin{center}
 \vspace{4cm}

 {\LARGE  Idealized Multigrid Algorithm for Staggered Fermions
          \footnote{Work supported by Deutsche Forschungsgemeinschaft.}}

 \addtocounter{footnote}{5}
 \vspace{3cm}
         Thomas Kalkreuter \footnote{Electronic address:
                                     i02kal@dsyibm.desy.de}\\
         \smallskip
    {\em II. Institut f\"ur Theoretische Physik
         der Universit\"at Hamburg, \\
         Luruper Chaussee 149,
         W-2000 Hamburg 50, Germany}

 \vspace{1.5cm}
 \mbox{} April 7, 1993

 \end{center}

 \vfill

 \begin{abstract}
 An idealized multigrid algorithm for the computation of propagators
 of staggered fermions is investigated.
 Exemplified in four-dimensional $SU(2)$ gauge fields, it is shown
 that the idealized algorithm preserves criticality under coarsening.
 The same is not true when the coarse grid operator is defined by
 the Galerkin prescription.
 Relaxation times in computations of propagators are small, and
 critical slowing is strongly reduced (or eliminated) in the
 idealized algorithm.
 Unfortunately, this algorithm is not practical for production runs,
 but the investigations presented here answer important questions
 of principle.
 \end{abstract}

 \vfill\mbox{}

 \renewcommand{\thefootnote}{\arabic{footnote})}
 \setcounter{footnote}{0}
  \newpage

 \section{Introduction}

 In Monte Carlo simulations of lattice gauge theories with fermions
 the most time-consuming part is the computation of the gauge field
 dependent fermion propagators.
 Great hopes to compute propagators without any critical slowing down
 (CSD) are attached to multigrid (MG) methods
 \cite{MGIsrael,SolLAT92,BRVGSP,MGBoston,MGAmsterdam,KalMacSpe,KalPL,%
       HHProceedings,KalLastPoint,KalLAT92,KalDiss}.
 However, up to now no practical MG algorithm has been found
 for fermions.

 In this note an idealized MG algorithm is investigated for staggered
 fermions in four-dimensional $SU(2)$ gauge fields.
 It will be shown that the idealized algorithm preserves criticality
 under coarsening, which is not true when the coarse grid operator is
 defined by the Galerkin prescription.
 This finding explains the failure of simple variational-like MG
 methods, at least for algorithms with nonoverlapping blocks or
 trivially overlapping blocks.

 Relaxation times in computations of propagators with the idealized
 MG algorithm are small, and CSD is strongly reduced or eliminated.
 Unfortunately, this algorithm is not practical for production runs,
 but the investigations presented here answer important questions of
 principle.

 \section{Multigrid Method}

 For given $f$ we wish to solve an equation
 \begin{equation}
   D_0\,\chi = f \enspace\mbox{with}\enspace
   D_0=( -\Dirac^2 + m^2 )
 \label{propagator}
 \end{equation}
 by MG, where $\Dirac$ is the gauge covariant staggered Dirac operator,
 and $m$ is a small quark mass.

 The following MG notations will be used.
 The fundamental lattice is denoted by $\Lambda^0$.
 The first block lattice $\Lambda^1$ is obtained by coarsening with a
 factor of $L_b$\@.
 Thus $\Lambda^1$ has $L_b^d$ fewer sites than $\Lambda^0$ (in $d$
 space-time dimensions).
 Restriction and interpolation operators $C$ and $\A$\/, respectively,
 are given by kernels $C(x,z)$ and $\A (z,x)$ with $z \in \Lambda^0$,
 $x \in \Lambda^1$\@.
 Note that $C(x,z)$ and $\A (z,x)$ are $\Nc\times\Nc$ matrices
 in a gauge theory with $\Nc$ colors.
 Also, $C$ and $\A$ depend on the gauge field, although this is not
 indicated explicitly.

 We use a blocking procedure for staggered fermions which is consistent
 with the lattice symmetries of free fermions \cite{KalMacSpe}.
 This forces us to choose $L_b =3$.
 Even $L_b$ are not allowed.
 In four dimensions, coarsening by a factor of three reduces the number
 of points by 81.
 Therefore only a two-grid algorithm was implemented.
 The residual equation on the coarse grid was solved exactly by
 the conjugate gradient algorithm.

 The averaging kernel $C$ is chosen according to the ground-state
 projection definition \cite{MacUnpublished,HSVGSP,BRVGSP,KalNP2}.
 In the present work $C$ fulfills the gauge covariant eigenvalue
 equation(s)
  \begin{equation}
  ( -\Delta_{N,x}\Cstar ) ( z , x )  = \lambda_{0}(x)\,\Cstar (z,x)
  \label{EVequationCSF}
  \end{equation}
 together with a normalization condition $C\Cstar = \bbbone$,
 and a covariance condition $C(x,\hat{x}) \propto \bbbone$
 where $\hat{x}$ denotes the center of block $x$.
 In Eq. (\ref{EVequationCSF}), $\lambda_0(x)$ is the lowest
 eigenvalue of $-\Delta_{N,x}$, and $-\Delta_{N,x}$ is the gauge
 covariant fermionic ``two-link lattice Laplacian'' -- defined through
 $\Dirac^2 = \Delta +
 \sigma_{\mu\nu}\,F_{\mu\nu}$ -- with ``Neumann boundary conditions
 (b.c.)''.
 Neumann b.c.\ means that terms in $\Delta$ are omitted where one
 site is in block $x$ and the other one is in a neighboring block.

 The ground-state projection method is numerically implementable
 in four-dimensional non-abelian gauge fields \cite{KalNP2}, and since
 the method is gauge covariant, no gauge fixing in computations of
 propagators is required.
 For staggered fermions in non-abelian gauge fields two qualitatively
 different proposals were made for ground-state projection
 \cite{KalMacSpe}.
 We call these proposals ``the Laplacian choice'' and ``the Diracian
 choice''.
 The Laplacian choice is the one described above.
 In the Diracian choice one substitutes a block-local approximation
 of $\Dirac^2$ for $\Delta$ in (\ref{EVequationCSF}).
 This latter choice would be superior because it takes also the
 field strength term $F_{\mu\nu}$ into consideration.
 However, it was proved numerically \cite{KalMacSpe,KalLAT92} that the
 Laplacian choice for $C$ defines a good blockspin in arbitrarily
 disordered gauge fields.%
 \footnote{For every averaging kernel $C$ there exists an associated
           ideal interpolation kernel $\A$; see Sec.~\ref{SecIdealMG}.
           $C$ defines a good blockspin if this $\A$ decays
           exponentially.}
 For this reason only the Laplacian choice of $C$ has been
 implemented yet.

 \section{Spectrum of $-\Dirac^2$ and CSD}\label{SecSpectrum}

 The square of the staggered Dirac operator (plus mass term) couples
 only even lattice sites to even sites, and odd sites to odd sites.
 Therefore the matrix elements of $-\Dirac^2$ can be arranged in such
 a way that $-\Dirac^2$ can be written symbolically as
 \begin{equation}
     -\Dirac^2  = \left( \begin{array}{cc}
                         -\Dirac^2_{\subs{even}}  & 0 \\
                         0 & -\Dirac^2_{\subs{odd}}   \\
                         \end{array} \right) \esk
 \label{decoupling}
 \end{equation}
 where $-\Dirac^2_{\subs{even/odd}}$ is $-\Dirac^2$ restricted to
 the even/odd sublattice.

 Let us denote the spectrum of $-\Dirac^2$ by $\S (-\Dirac^2 )$.
 It equals the union of the spectra of $-\Dirac^2_{\subs{even}}$ and
 $-\Dirac^2_{\subs{odd}}$:
 $\S (-\Dirac^2 ) = \S (-\Dirac^2_{\subs{even}} ) \bigcup
                    \S (-\Dirac^2_{\subs{odd}} )$.
 The spectra are gauge invariant.
 Moreover, for any gauge field configuration one has the equality
 \begin{equation}
  \S (-\Dirac^2_{\subs{even}} )  =
  \S (-\Dirac^2_{\subs{odd}} ) \esp
 \label{IdenticalSpectra}
 \end{equation}
 A simple proof of (\ref{IdenticalSpectra}) is as follows.
 Consider the lattice operator $-\Dirac^2_{\subs{even/odd}}$ as a
 block matrix with $\Nc\times\Nc$ elements
 $-\Dirac^2_{\subs{even/odd}}(z_1,z_2)$.
 The matrices $-\Dirac^2_{\subs{even}}$
 and $-\Dirac^2_{\subs{odd}}$
 are similar, and therefore they have the same spectrum.
 We recall that two matrices $A$ and $B$ are called similar if
 there exists an invertible matrix $T$ such that $B = T A T^{-1}$.
 (We also recall that if $v$ is an eigenvector of $A$ with eigenvalue
  $\lambda$, then $Tv$ is an eigenvector of $B$ with eigenvalue
  $\lambda$.)
 In the case considered here, we look for a lattice operator $T$
 with the property
 \begin{equation}
  \sum_{z'} T (w,z') (-\Dirac^2_{\subs{even}})(z',z)  =
  \sum_{w'} (-\Dirac^2_{\subs{odd}})(w,w') T (w',z)
 \label{T}
 \end{equation}
 where $z$, $z'$ and $w$, $w'$ denote even and odd lattice sites,
 respectively.
 Eq.~(\ref{T}) is fulfilled if we choose the matrix elements of $T$
 to be $T(w,z) = \Dirac (w,z)$.
 This choice of $T$ is not invertible in pure gauges, but in that case
 the equality (\ref{IdenticalSpectra}) of spectra is obvious anyhow.

 In Ref.~\cite{MarOtt} a more complicated proof of
 (\ref{IdenticalSpectra}) was given which uses an analyticity
 argument in connection with a hopping expansion of
 $(-\Dirac^2 + m^2 )^{-1}$ for large mass.

 In Refs.~\cite{KalLastPoint,KalLAT92,KalDiss} the author pointed
 out that in conventional relaxation algorithms for propagators
 of staggered fermions there exists a scaling law for relaxation times
 $\tau$ which reads
 \begin{equation}
   \tau = \frac{\mbox{{\em const.}}}{\Dm} \quad \mbox{with} \quad
   \Dm = m^2 - \mcr
 \label{ScalingTau}
 \end{equation}
 for small $\Dm$,
 where $\mcr$ is the lowest eigenvalue of $-\Dirac^2$, and {\em const.}
 is {\em independent of the lattice size\/}.
 For bosonic propagators the validity of (\ref{ScalingTau}) is known
 analytically \cite{KalDiss} and has also been confirmed to a high
 accuracy numerically \cite{KalPL,KalDiss}.

 A consequence of (\ref{IdenticalSpectra}) is that in conventional
 relaxation algorithms for staggered fermions, CSD will be the same
 on the even and the odd sublattice.

 \section{Idealized Multigrid Algorithm}\label{SecIdealMG}

 Up to now no practical MG method has been found for fermions.
 Mack pointed out that it is essential for fighting CSD in MG
 computations that interpolation kernels should be smooth
 \cite{MacCargese}.
 This requirement is not fulfilled in MG algorithms where one uses
 gauge covariant generalizations of piecewise constant interpolation
 with nonoverlapping blocks.
 Mack suggested an interpolation kernel $\A$ as a starting point for
 numerical work which was used successfully by Gaw\c{e}dzki and
 Kupiainen in constructive quantum field theory \cite{GawKup}.
 An idealized MG algorithm using the natural gauge covariant
 generalization of the Gaw\c{e}dzki-Kupiainen kernel had been
 investigated numerically in four-dimensional $SU(2)$ gauge fields
 for bosonic propagators \cite{KalPL,KalDiss}.
 There CSD could be eliminated completely.
 Here we turn to an idealized MG algorithm for staggered fermions.
 This algorithm will not be practical for production runs, but it is
 important to answer questions of principle, and to recognize the
 features which a successful method must have.

 Given the averaging kernel $C$, there exists an ideal choice
 of the interpolation kernel $\A$.
 It is determined as follows.
 For every function (``block spin'') $\Phi$ on $\Lambda^1$,
 $\phi = \A \Phi$ minimizes the action $< \phi\,,\,D_0\,\phi >$
 subject to the constraint $C \phi = \Phi$.
 For the purpose of numerical computations, it is convenient to
 determine the optimal $\A$ as the solution of the equation
 \begin{equation}
  \left( [ -\Dirac^2 + m^2 + \kappa\,\Cstar C ] \A \right) (z,x)
  = \kappa\,\Cstar (z,x)
 \label{optimalA}
 \end{equation}
 for large $\kappa$.
 $\Cstar$ denotes the adjoint of $C$\@.
 The layers of an MG decouple completely when this $\A$ is used for
 interpolation, and when coarse grid operators $D_1$ are defined as
 $C ( -\Dirac^2 + m^2 ) \A$.
 These coarse grid operators are automatically hermitean and equal
 $\A^{\ast} ( -\Dirac^2 + m^2 ) \A$.

 The optimal $\A$ is favored by an argument of dynamics
 \cite{HHProceedings,KalDiss}.
 With the definition of smoothness that covariant derivatives are
 small, the above characterization of $\A$ as solution of an
 extremization problem can be rephrased:
 that $\A$ is the smoothest interpolation kernel, subject to the
 constraint $C\A = \bbbone$.

 \section{Numerical Results}

 Because of the considerations of Sec.~\ref{SecSpectrum} we
 decided to investigate the idealized two-grid algorithm only
 on the even sublattice.
 This restriction mitigates also the storage space requirements for the
 ideal $\A$ a little bit.
 Note that Eq.~(\ref{optimalA}) can also be broken up into an
 equation for the interpolation kernel on the even sublattice and
 one on the odd one, if the averaging kernel $C$ does not mix
 even and odd sites.
 This requirement is fulfilled both for the Laplacian and for the
 Diracian choice of Ref.~\cite{KalMacSpe}.
 In both proposals the coarse grid sites can be separated into
 even and odd sites, and the ideal effective Diracian $-C\Dirac^2\A$
 (as well as the Galerkin operator $-C\Dirac^2\Cstar$)
 can be decomposed analogously to (\ref{decoupling}).

 Here we made the Laplacian choice and computed $C$ by the efficient
 algorithm of Ref.~\cite{KalNP2}.
 Numerical work was done in $SU(2)$ lattice gauge fields on $6^4$ and
 $12^4$ lattices, covering all possible values of $\beta = 4/g^2$
 between $\infty$ and zero.
 The system (\ref{optimalA}) was solved by means of the conjugate
 gradient algorithm where iterating was stopped when the RMS residual
 was less than $10^{-10}$.
 The statement that $-C\Dirac^2\A$ is automatically hermitean
 was confirmed up to round-off errors of order $10^{-9}$ or less.

 \subsection{Lowest Eigenvalues}

 Let us first look at the lowest eigenvalues of $-\Dirac^2$
 on the fundamental lattice $\Lambda^0$, and see how they are
 transferred to the block lattice $\Lambda^1$\@.
 The role of this transfer for the performance of the
 parallel-transported MG method of Ben-Av, Brandt and Solomon was
 pointed out in Ref.~\cite{SolLAT92}.

 First the lowest eigenvalues $-\mcr$ of $-\Dirac^2$ were determined
 by inverse iteration.
 This method allows to determine $-\mcr$ to an accuracy of $10^{-7}$
 or better \cite{KalPL}.
 Then optimal interpolation kernels $\A$ were computed as solutions
 of Eq.~(\ref{optimalA}) with $m^2 = \mcr$\,, and for $\kappa = 10^{5}$.
 Results for $-\mcr$ and for the lowest eigenvalues of the
 ideal coarse grid operator $C (-\Dirac^2 + \mcr ) \A$ are given
 in Table \ref{TableEigenvalues}.
 The last column of  Table \ref{TableEigenvalues} contains results
 for the Galerkin definition of the coarse grid operator.
 This operator is used in variational MG where interpolation is done
 by $\Cstar$.
 (The Galerkin operator retains the locality properties of $-\Dirac^2$
 in arbitrary gauge fields.)

 One sees that for any value of the gauge coupling the idealized
 algorithm maps a critical system on $\Lambda^0$ onto a critical system
 on the block lattice.
 In contrast, variational MG does not have this property.
 The Galerkin operator $C (-\Dirac^2 + \mcr ) \Cstar$ is far from
 being critical in nontrivial gauge fields.

 The results of Table \ref{TableEigenvalues} supply another
 explanation for the failure of variational MG which was ascertained
 in Refs.~\cite{KalLastPoint,KalLAT92,KalDiss}.
 One cannot expect that a (nearly) critical problem on $\Lambda^0$
 can be solved by means of an auxiliary problem with fewer degrees
 of freedom on $\Lambda^1$, if the auxiliary problem is not critical
 as well.
 The Galerkin operator is only critical in trivial gauge fields,%
 \footnote{In exact arithmetics the entries for $\beta = \infty$
           (realized as random pure gauge fields) in Table
           \ref{TableEigenvalues} would be zero for any lattice size.}
 and only there is CSD eliminated by the variational MG method
 in computations of propagators.

 The effect of adding a mass term $\Dm$ to
 $(-\Dirac^2 + \mcr )$ is as follows.
 The values given in the last two columns of Table
 \ref{TableEigenvalues} are shifted by the amount determined by
 $\Dm$.
 This is obvious in case of the Galerkin operator because the
 averaging kernel is normalized such that $C \Cstar = \bbbone$.
 In case of the idealized algorithm, $C\A$ tends to $\bbbone$
 for $\kappa\rightarrow\infty$.
 For finite $\kappa$ one finds deviations from $\bbbone$.
 Examples are given in Table \ref{TableDeviation}.
 These deviations are small enough to have no effect in practice.
 When one computes the lowest eigenvalues of $-C\Dirac^2\A$,
 one recovers the negative critical masses of the gauge fields to the
 same accuracy as they are given in Table \ref{TableEigenvalues}.
 Hence, the small negative values partly found for the lowest
 eigenvalues of $C (-\Dirac^2 + \mcr ) \A$ are really due to numerical
 inaccuracies.

 \subsection{Performance of the idealized algorithm}

 Finally, we report results of computations of propagators by means
 of an idealized algorithm.
 We used coarse grid operators $C (-\Dirac^2 + m^2)\A$
 with masses $m^2 = \mcr + \Dm$, $\Dm > 0$ and small.
 For all values of $\Dm$ only one interpolation kernel $\A$ was used,
 namely the ``critical'' one which solves (\ref{optimalA}) with $m^2 =
 \mcr$.
 Actually, one should use an $m^2$-dependent $\A$-kernel, viz.\
 the solution of (\ref{optimalA}) with $m^2$ being the mass under
 consideration.
 However, in case of bosonic propagators the procedure described
 here was successfull \cite{KalPL,KalDiss}, and therefore we used
 it also as a first attempt in case of staggered fermions.

 Tables \ref{Table6**4} and \ref{Table12**4} comprise results
 for relaxation times as a function of $\Dm$ on $6^4$ and $12^4$
 lattices, respectively.
 The values given for the relaxation parameter $\omega$ are optimal
 within $\pm 0.05$.
 (For the configuration on the $12^4$ lattice at $\beta = 2.5$, $\omega
  = 1.72$ is close to optimum, while the results for $\omega = 1.65$
  are given as additional information.)

 We rediscover here an observation which was made earlier
 \cite{KalPL,KalDiss}:
 The use of a relaxation parameter $\omega$ different from one
 in MG computations contradicts the conventional wisdom.
 According to this wisdom the only job of the relaxation procedure
 on $\Lambda^0$ is to smoothen the error, and this job is well done
 by Gauss-Seidel iteration.
 The conventional wisdom was confirmed by numerical results in trivial
 gauge fields \cite{KalDiss}.
 However, the picture changes for propagators in nontrivial gauge
 fields.

 In order to be sure about the correct determination of the value of
 $\mcr$, it was checked that conventional SOR and the variational MG
 algorithm (with the Laplacian choice of $C$) both exhibited CSD, i.e.
 relaxation times $\tau$ follow perfectly the scaling law
 (\ref{ScalingTau}).
 $\mbox{const.}$ in (\ref{ScalingTau}) is of order one.

 The results of Tables \ref{Table6**4} and \ref{Table12**4} show that
 the $1/\Dm$ divergence of $\tau$'s on lattices of a fixed size is
 eliminated in the idealized MG algorithm.
 Relaxation times are bounded and small.
 (Only for very disordered gauge fields at physically uninteresting
 values of the gauge coupling, $\tau$'s are not so small, but
 nevertheless bounded.)
 We conclude from Tables \ref{Table6**4} and \ref{Table12**4} that
 CSD in computation of propagators is strongly reduced.
 It is hard to judge a possibly remaining volume effect, but one
 might be tempted to say that CSD can be eliminated for practical
 purposes, in principle.
 At this point one should also recall that in the investigations
 reported here, only one $\A$-kernel was used for all values of $m^2$.
 If one used an $m^2$-dependent kernel, the results would probably
 improve further, at least they cannot become worse.

 \smallskip\noindent{\sc Acknowledgments}\newline\indent
 I wish to thank Gerhard Mack for many stimulating discussions.
 Financial support by Deutsche Forschungsgemeinschaft is gratefully
 acknowledged.
 The computations reported here were performed on the CRAY Y-MP
 of HLRZ J\"ulich.


 \newpage
 \section*{Tables}
 \renewcommand{\arraystretch}{1.2}
 \begin{table}[h]
 \caption{Lowest eigenvalues of the negative squared Dirac operator
          for staggered fermions on a fundamental lattice $\Lambda^0$,
          and the lowest eigenvalues of the ideal block
          operator $C(-\Dirac^2+\protect\mcr )\A$ and of the
          Galerkin operator $C (-\Dirac^2 + \protect\mcr )\Cstar$.}
 \label{TableEigenvalues}
 \begin{minipage}{17 cm}{}
 \begin{center}
 \begin{tabular}{ccccc}
 \hline
         &          & $-\protect\mcr =$ lowest eigenvalue of
                    &lowest eigenvalue of & lowest eigenvalue of \\
 $\beta$ & $| \Lambda^0|$ & $-\Dirac^2$ & $C(-\Dirac^2+\protect\mcr )\A$
                               & $C (-\Dirac^2 + \protect\mcr )\Cstar$\\
 \hline
 $\infty$ &  $6^4$ &  7.98$\cdot 10^{-29}$ &\p 3.09$\cdot 10^{-21}$
                   &  9.04$\cdot 10^{-13}$ \\
 $5.0$    &  $6^4$ & 0.0013413 & $-$1.40$\cdot 10^{-11}$ & 1.0341375\\
 $3.0$    &  $6^4$ & 0.3497739 &\p  6.68$\cdot 10^{-12}$ & 2.8529730\\
 $2.8$    &  $6^4$ & 0.2441995 &\p  7.27$\cdot 10^{-11}$ & 2.8412484\\
 $2.8$    &  $6^4$ & 0.2748178 &\p  4.33$\cdot 10^{-12}$ & 3.2338422\\
 $2.7$    &  $6^4$ & 0.2004647 & $-$6.42$\cdot 10^{-11}$ & 3.4374150\\
 $2.6$    &  $6^4$ & 0.1740946 & $-$2.78$\cdot 10^{-11}$ & 3.1884409\\
 $2.5$    &  $6^4$ & 0.0698942 & $-$5.96$\cdot 10^{-13}$ & 3.0198011\\
 $2.4$    &  $6^4$ & 0.0010729 & $-$1.82$\cdot 10^{-11}$ & 3.1265394\\
 $2.2$    &  $6^4$ & 0.0007099 &\p  9.50$\cdot 10^{-11}$ & 3.4488349\\
 $2.0$    &  $6^4$ & 0.0000732 &\p  6.08$\cdot 10^{-10}$ & 3.5620285\\
 $0.0$    &  $6^4$ & 0.0000287 & $-$1.21$\cdot 10^{-10}$ & 3.7354123\\
 $\infty$ & $12^4$ &  8.15$\cdot 10^{-29}$ & $-$2.33$\cdot 10^{-14}$
                   &  9.13$\cdot 10^{-13}$ \\
 $3.0$    & $12^4$ & 0.0779810 & $-$7.96$\cdot 10^{-12}$ & 2.3190549\\
 $2.7$    & $12^4$ & 0.0368447 & $-$1.15$\cdot 10^{-11}$ & 2.3967694\\
 $2.5$    & $12^4$ & 0.0005742 & $-$1.03$\cdot 10^{-10}$ & 2.6881098\\
 $2.4$    & $12^4$ & 0.0001865 & $-$1.10$\cdot 10^{-8\pn}$ &2.8248637\\
 \hline
 \end{tabular}
 \end{center}
 \end{minipage}
 \end{table}

 \begin{table}
 \caption{Accuracy of $C \A = \protect\bbbone$ for staggered
          fer\-mi\-ons.
          $\| C \A - \protect\bbbone \|_{\infty}$ denotes the maximal
          trace norm of $C \A (x,y) - \delta (x-y)$ over all pairs
          $(x,y)$ of block lattice sites, and
          $\| C \A - \protect\bbbone \|_{2}$ is the RMS of these
          norms.}
 \label{TableDeviation}
 \begin{minipage}{17 cm}{}
 \begin{center}
 \begin{tabular}{ccll}
 \hline
 $\beta$ & $| \Lambda^0 |$ & $\| C \A - \protect\bbbone \|_2$
                         & $\| C \A - \protect\bbbone \|_{\infty}$ \\
 \hline
 $\infty$%
 \footnote{In case of the pure gauge on the $6^4$ lattice (and only in
            this case!), the ideal MG scheme is identical to the
            Galerkin definition with $\A = \Cstar$ (``covariant
            piecewise constant'' interpolation);
            since $C$ is normalized as $C\Cstar = \protect\bbbone$,
            the finite norms of $C \A - \bbbone$ in this case are due
            to round-off errors.}
          &  $6^4$ & 3.98$\cdot 10^{-15}$& 2.13$\cdot 10^{-14}$ \\
 $5.0$    &  $6^4$ & 6.62$\cdot 10^{-7}$ & 1.87$\cdot 10^{-6}$ \\
 $3.0$    &  $6^4$ & 2.77$\cdot 10^{-6}$ & 9.43$\cdot 10^{-6}$ \\
 $2.8$    &  $6^4$ & 2.59$\cdot 10^{-6}$ & 8.66$\cdot 10^{-6}$ \\
 $2.8$    &  $6^4$ & 3.69$\cdot 10^{-6}$ & 1.07$\cdot 10^{-5}$ \\
 $2.7$    &  $6^4$ & 3.76$\cdot 10^{-6}$ & 1.18$\cdot 10^{-5}$ \\
 $2.6$    &  $6^4$ & 3.18$\cdot 10^{-6}$ & 1.02$\cdot 10^{-5}$ \\
 $2.5$    &  $6^4$ & 2.43$\cdot 10^{-6}$ & 8.15$\cdot 10^{-6}$ \\
 $2.4$    &  $6^4$ & 2.39$\cdot 10^{-6}$ & 6.76$\cdot 10^{-6}$ \\
 $2.2$    &  $6^4$ & 1.51$\cdot 10^{-6}$ & 4.48$\cdot 10^{-6}$ \\
 $2.0$    &  $6^4$ & 9.40$\cdot 10^{-7}$ & 2.95$\cdot 10^{-6}$ \\
 $0.0$    &  $6^4$ & 2.84$\cdot 10^{-7}$ & 9.23$\cdot 10^{-7}$ \\
 $\infty$ & $12^4$ & 2.32$\cdot 10^{-6}$ & 2.32$\cdot 10^{-5}$ \\
 $3.0$    & $12^4$ & 1.02$\cdot 10^{-6}$ & 1.12$\cdot 10^{-5}$ \\
 $2.7$    & $12^4$ & 1.00$\cdot 10^{-6}$ & 1.11$\cdot 10^{-5}$ \\
 $2.5$    & $12^4$ & 9.08$\cdot 10^{-7}$ & 1.05$\cdot 10^{-5}$ \\
 $2.4$    & $12^4$ & 7.92$\cdot 10^{-7}$ & 9.12$\cdot 10^{-5}$ \\
 \hline
 \end{tabular}
 \end{center}
 \end{minipage}
 \end{table}

 \begin{table}
 \caption{Dependence of relaxation times $\tau$ on $\Dm$ in the
          ideal MG algorithm with lexicographic SOR for computing
          propagators of staggered fermions on $6^4$ lattices.
          (The two configurations at $\beta = 2.8$ are different.)}
 \label{Table6**4}
 \begin{minipage}{17 cm}{}
 \begin{center}
 \begin{tabular}{ccrrrrrrr}
 \hline
         &          & \multicolumn{7}{c}{$\tau\mbox{\ for\ }\Dm =$} \\
 $\beta$ & $\omega$ & $1$ & $10^{-1}$ & $10^{-2}$ & $10^{-3}$
                          & $10^{-4}$ & $10^{-5}$ & $10^{-6}$ \\
 \hline
 $\infty$
       & 1.17 & 0.8 & 0.8 & 0.7 & 0.7 & 0.7 & 0.7 & 0.7 \\
 $5.0$ & 1.35 & 1.2 & 1.3 & 1.4 & 1.4 & 1.4 & 1.4 & 1.4 \\
 $3.0$ & 1.70 & 4.0 & 4.5 & 4.8 & 4.9 & 4.9 & 4.9 & 4.9 \\
 $2.8$ & 1.70 & 3.6 & 4.4 & 5.4 & 5.8 & 5.9 & 5.9 & 5.9 \\
 $2.8$ & 1.75 & 5.2 & 6.6 & 7.1 & 7.7 & 7.8 & 7.8 & 7.8 \\
 $2.7$ & 1.65 & 3.5 & 4.5 & 5.0 & 6.1 & 6.2 & 6.2 & 6.2 \\
 $2.6$ & 1.65 & 3.2 & 4.5 & 5.2 & 6.0 & 6.1 & 6.1 & 6.0 \\
 $2.5$ & 1.65 & 3.2 & 4.4 & 6.8 & 7.5 & 7.6 & 7.6 & 7.6 \\
 $2.4$ & 1.72 & 4.5 & 7.2 & 12. & 16. & 17. & 17. & 17. \\
 $2.2$ & 1.70 & 6.2 & 15. & 20. & 43. & 53. & 54. & 80. \\
 $2.0$ & 1.60 & 5.0 & 13. & 54. & 139.& 199.& 211. & 213. \\
 $0.0$ & 1.45 & 28. & 36. & 119.& 524.& 1156.& 1505. & 1570. \\
 \hline
 \end{tabular}
 \end{center}
 \end{minipage}
 \end{table}

 \begin{table}
 \caption{Dependence of relaxation times $\tau$ on $\Dm$ in the
          ideal MG algorithm with lexicographic SOR for computing
          propagators of staggered fermions on $12^4$ lattices.
          (The values given for $\beta = 2.5$ were obtained in the same
           gauge field configuration.)}
 \label{Table12**4}
 \begin{minipage}{17 cm}{}
 \begin{center}
 \begin{tabular}{ccrrrrrrr}
 \hline
         &          & \multicolumn{7}{c}{$\tau\mbox{\ for\ }\Dm =$} \\
 $\beta$ & $\omega$ & $1$ & $10^{-1}$ & $10^{-2}$ & $10^{-3}$
                          & $10^{-4}$ & $10^{-5}$ & $10^{-6}$ \\
 \hline
 $\infty$ & 1.32 & 1.5 & 1.7 & 1.7 & 1.7 & 1.7 & 1.7 & 1.7 \\
 $3.0$ & 1.65 & 2.9 & 3.2 & 3.3 & 3.5 & 3.5 & 3.5 & 3.5 \\
 $2.7$ & 1.65 & 2.9 & 3.4 & 4.0 & 4.5 & 4.5 & 4.5 & 4.5 \\
 $2.5$ & 1.72 & 4.2 & 4.9 & 6.9 & 8.4 & 8.7 & 8.8 & 8.8 \\
 $2.5$ & 1.65 & 3.2 & 3.8 & 6.9 & 8.8 & 9.3 & 9.4 & 9.4 \\
 $2.4$ & 1.65 & 3.3 & 5.0 & 15.4 & 23.6 & 27.7 & 28.0 & 28.0\\
 \hline
 \end{tabular}
 \end{center}
 \end{minipage}
 \end{table}

\end{document}